# Designing The Internet of Agents: A Framework for Trustworthy, Transparent, and Collaborative Human-Agent Interaction (HAX)

| Marc Scibelli | Krystelle Gonzalez Papaux | Julia Valenti | Srishti Kush |
|---|---|---|---|
| Outshift by Cisco | University of Technology of Compiegne, Outshift by Cisco | Outshift by Cisco | Outshift by Cisco |

## Abstract

*The rise of generative and autonomous agents marks a fundamental shift in computing, demanding a rethinking of how humans collaborate with probabilistic, partially autonomous systems. We present the Human-AI-Experience (HAX) framework, a comprehensive, three-phase approach that establishes design foundations for trustworthy, transparent, and collaborative agentic interaction. HAX integrates behavioral heuristics, a schema-driven SDK enforcing structured and safe outputs, and a behavioral proxy concept that orchestrates agent activity to reduce cognitive load. A validated catalog of mixed-initiative design patterns further enables intent preview, iterative alignment, trust repair, and multi-agent narrative coherence. Grounded in Time, Interaction, and Performance (TIP) theory, HAX reframes multi-agent systems as colleagues, offering the first end-to-end framework that bridges trust theory, interface design, and infrastructure for the emerging Internet of Agents.*

## 1. Introduction

The emergence of generative and autonomous agents represents a transformative shift in the trajectory of computing, fundamentally challenging established paradigms of interface design. As these systems transition from static tools to adaptive collaborators, the imperative for novel design frameworks becomes increasingly evident. Such agents are no longer confined to executing predefined tasks; they are expected to operate in concert with human users, interpret situational context, make independent decisions, and initiate actions autonomously. Currently, prevailing models of user interface design remain ill-equipped to address the corresponding demands for transparency, error recovery, and the negotiation of shared control that these systems require to obtain optimal user experiences. Traditional design frameworks were conceived for systems with predictable behaviors, stable affordances, and clearly bounded states, where the user could maintain a coherent mental model of system operation. In contrast, autonomous agents act with degrees of independence, uncertainty, and interpretive variability that resists such predictability. The resulting challenge for user experience is threefold.

First, transparency can no longer be achieved solely through visual cues or explicit feedback; it requires dynamic representations of the agent's reasoning process, intent, and evolving internal state elements that are often opaque or distributed across multiple interacting agents.

Second, error recovery in these contexts cannot rely on linear undo mechanisms or static fail-safes. When actions emerge from probabilistic or generative reasoning, recovery must involve interpretive dialogue, a process of diagnosing misalignments, revising shared goals, and collaboratively redefining acceptable system behaviors.

Third, the negotiation of shared control introduces new tensions between human authority and machine autonomy. Interfaces must support fluid transitions of initiative, enabling users to delegate, override, or co-steer actions in ways that preserve both human oversight and agentic responsiveness. Yet existing UI paradigms—rooted in command hierarchies or deterministic control metaphors, lack the representational and interactional scaffolds needed for such reciprocal coordination.

This paper introduces the HAX framework, a three-phase initiative that sheds light on how best to address these changes and create the foundation for trust, transparency, collaboration, and accountable, interpretable coordination between humans and agents in this new paradigm.

It comprises a set of design heuristics (Phase 1), a structured SDK for agentic content (Phase 2), and a proposed behavioral orchestration agent (Phase 3). Together, these phases provide developers, designers, and researchers with practical tools and behavioral models for building interfaces that are trustworthy, transparent, and collaborative.

We further present a set of design principles and an evolving component catalog derived from real-world use cases and informed by grounded theory research, thematic and ethnographic analysis, including insights from academic literature and peer-reviewed studies. These resources aim to address the persistent gap between agentic capabilities and human trust[1] providing structured approaches for designing interactions in which humans and agents engage in co-creative processes rather than unidirectional command–execution dynamics.

## 2. Background and Motivation

The trajectory of computing has long been shaped by the metaphors of control and command. From early graphical user interfaces to today's conversational agents, design paradigms have primarily evolved to make increasingly complex systems legible and manipulable to human users. However, these paradigms assumed a fundamental asymmetry: the human as sole initiator and decision-maker, and the machine as executor of predefined operations.

The rise of machine learning, generative AI, and multi-agent architectures disrupts this foundational assumption. Systems now exhibit behaviors that are context-dependent, stochastic, and emergent, often producing outcomes that cannot be fully anticipated by their creators or users. As these systems acquire the capacity to interpret goals, generate strategies, and act autonomously, interaction design must grapple with a new form of agency, one that is coextensive with human cognition rather than subordinate to it.

This shift exposes new challenges within the field of Human–Computer Interaction (HCI). Traditional interface frameworks emphasize efficiency and predictability and were rooted primarily in usability. They were designed around systems whose behavior could be modeled, optimized, and predicted, privileging efficiency, consistency, and error-free execution as the central criteria of interface quality.[2] Yet interactions with autonomous agents hinge instead on interpretation, trust, and negotiation and require sense-making, trust calibration, and ongoing negotiation of intent, since such systems exhibit adaptive and emergent behaviors that cannot be fully specified or anticipated in advance.[3]

Despite the rapid advancement of AI's functional capacities, the design of user interfaces has yet to keep pace in supporting this new type of collaborative and agentic form of interaction. Although establishing "appropriate trust" is widely recognized as essential for effective human–AI interaction[4], the field remains fragmented and lacks a shared definition on what this entails. Scholars disagree on how to distinguish between related notions such as trust, reliance, and trustworthiness, creating conceptual ambiguity.[5]

Researchers and designers have proposed multiple strategies to foster appropriate trust in AI systems, including confidence scores [6] [7] interpretability methods [8], and signaling mechanisms such as alarms [9], warning indicators [10], or uncertainty communication.[11] Much of these works seeks to align user trust with the actual trustworthiness of the system. Yet, the field still lacks a unified framework, and the concept of "appropriate trust" remains contested.[12] As such, when AI agents operate with partial autonomy, opaque internal logic, and dynamically evolving memory systems, end users often have trouble in predicting system behaviors and outcomes, comprehending agent reasoning, or intervening appropriately when needed.

This conceptual and practical fragmentation is mirrored in the developer ecosystem. Developers today create bespoke agentic programs for each use case, writing custom prompts, custom chat interfaces, and ad hoc tool integrations, resulting in inconsistent experiences, duplicated efforts, and a lack of shared visual and behavioral patterns.

Without standardized frameworks for designing, observing, and governing agentic interactions, each implementation reinvents basic interaction primitives, further exacerbating the opacity and unreliability of multi-agent workflows. What emerges is a landscape of isolated prototypes rather than cohesive, interoperable design conventions that could sustain trust, legibility, and be reused across agentic systems. The absence of shared design standards for multi-agent interaction is not merely a tooling inconvenience—it produces opaque and non-generalizable orchestration practices that directly compromise interpretability and end-user control. Because each system constructs its own orchestration metaphors, monitoring channels, and intervention rules, users must continually relearn how to trace decisions, observe agent negotiations, and identify where failure originates. Without conventions for transparency, hierarchy visualization, or inter-agent communication exposure, users are forced into roles of accidental system supervisors—expected to manage cascading failures or hidden delegation chains without the interpretive tools to understand how agents interact or why outcomes emerge, significantly impacting user trust and increasing cognitive load.

In multi-agent systems, multiple agents may act concurrently, offer conflicting advice, or act in parallel. Without a coherent UI framework, these systems become noisy, unpredictable, and difficult to manage.

**Example**

*In hierarchical architectures in which a single orchestrator delegates tasks to specialized sub-agents such as a travel-planning system - a Booking Agent may manage reservations, a Budget Agent may be responsible for modifying costs, and a Calendar Agent will schedule meetings—each acting semi-autonomously beneath the surface. A user requesting a simple 5-day travel itinerary may see the Booking Agent secure a hotel, the Budget Agent immediately cancel and rebook cheaper accommodations, and the Calendar Agent schedule commitments based on the now-invalid first reservation, all without any visible explanation or rollback mechanism. Trust is eroded not because the agents are incapable, but because users cannot verify behavior, recover from cascading mistakes, or understand how parallel decisions were made in the first place. Put simply, trust is thus broken when users cannot verify agent behavior, recover from mistakes, or understand system decisions.*[13]

Although much of agent coordination in multi-agent systems occur beyond the user's immediate perception, it nonetheless represents a fundamental interactional concern, insofar as its outcomes directly shape the user's experience of coherence, control, and trust. The experiential consequences of hidden agentic negotiations, manifesting as inconsistent outputs, ambiguous system states, or fluctuating response latencies, directly affect the user's capacity to form a stable mental model of the system.

Moreover, as multi-agent architectures are increasingly integrated into developer-facing, operational, and decision-support environments, users are not passive recipients of automated outcomes but rather co-participants in distributed cognition[14]. Their work involves inspecting, diagnosing, and at times overriding autonomous processes. Consequently, the interface becomes less a site of command and control and more of a cognitive mediation surface through which the reasoning and coordination of agents are rendered perceptible, interpretable, and alignable with human intent.

What is required is not merely the refinement of a given user interface design paradigm, but the establishment of a new behavioral protocol for human-agent collaboration. Such a protocol should embed principles of explainability, reversibility, collaboration, and predictability as core design requirements, ensuring that human–agent interaction remains transparent, accountable, and controllable.

## 3. The HAX Framework

The Human-AI-Experience (HAX) framework outlines a three-phase approach to redesigning interfaces for agentic interaction. While this paper focuses on the first two phases, the framework establishes a progression from design principles to technical enforcement and, ultimately, adaptive orchestration.

- **Phase 1: Design Heuristics**
  Defines opinionated principles for control, clarity, recovery, collaboration, and trust. These function not as interface components but as behavioral guardrails for agentic systems.
- **Phase 2: Agentic SDK**
  Introduces a schema-driven software development kit that enables agents to safely populate structured UI blocks. Developers retain control of layout and rendering logic, while agents supply structured intent.
- **Phase 3: The HAX Agent** (beyond the scope of this paper)
  A behavioral proxy that filters, sequences, and adapts agent outputs before they reach the interface. It personalizes presentation based on user context, reducing cognitive load and improving clarity.

Together, these phases build upon one another, from heuristics to enforcement to orchestration, supporting scalable and modular adoption across agentic product ecosystems.

## 4. Agentic Design Principles

As the first phase of the HAX Framework, the Agentic Design Principles define the foundational vocabulary for shaping human–agent collaboration. The five principles (Control, Clarity,

Recovery, Collaboration, and Traceability) serve as behavioral guardrails that guide the development of trustworthy, transparent, and mixed-initiative systems.

Each principle was iteratively developed through interviews, field studies, and validation cycles, grounding the framework in both empirical evidence and practitioner experience. We engaged with over 40 experts spanning multiple sectors and disciplinary domains—including leading scholars in artificial intelligence, human–computer interaction, and socio-technical systems, as well as practitioners from advanced technology, enterprise software, and large-scale computational infrastructure industries.

Additionally, we draw on Time, Interaction, and Performance (TIP) theory[15] which conceptualizes groups as temporal, multimodal, and multifunctional systems. TIP identifies three core functions, production (goal attainment), support (affective exchange), and well-being (maintenance of norms and roles), that sustain coordination and reliability over time.

Within this framework, McGrath defines four modes of group operation—Inception, Problem-Solving, Conflict Resolution, and Execution that describe how groups organize and adapt to shifting demands. These modes are not linear stages but recurrent patterns of activity that enable teams to initiate, deliberate, repair, and act in pursuit of shared objectives.

Building on this foundation, we extend TIP to the design of multi-agent systems as a framework for *trust calibration*. Rather than redefining these modes, we use them to determine which aspects of agentic activity should be made visible to human collaborators.

| TIP Mode | Primary HAX Principle(s) | What to Surface to User |
| --- | --- | --- |
| **Inception** | Control + Clarity | Reveal agent goals, planned actions, and assumptions before execution; enable user approval or adjustment. |
| **Problem Solving** | Collaboration + Clarity | Expose reasoning, trade-offs, and confidence levels while the agent evaluates options; invite user input or correction. |
| **Conflict Resolution** | Recovery + Collaboration | Surface disagreements, overrides, or detected anomalies; present clear paths to correction or explanation of divergent reasoning. |
| **Execution** | Traceability + Recovery | Show progress, completion status, and logs of decisions; allow rollback or verification of prior steps. |

Table 1. TIP Applicability to HAX Principles

Together, these principles form a shared design vocabulary for trustworthy agentic workflows, bridging theoretical models of appropriate trust with the practical realities of building usable, transparent, and accountable multi-agent systems.

The following examples and figures illustrate how each principle comes to life in interface patterns and visual treatments, translating abstract design goals into concrete forms of interaction.

1. **Control:** Control emphasizes how humans shape and constrain agent behavior by setting boundaries, preferences, and permissions. Rather than limiting autonomy, these mechanisms align agent initiatives with human intent, enabling flexible delegation and oversight.

    UI patterns include scope and boundary settings, customization of autonomy, and permission gates, which give users structured means to negotiate shared control while maintaining situational authority.

    Figure 1 demonstrates this: users define interaction boundaries by selecting input modes, indicating what the agent is allowed to modify and what it must avoid. This establishes clear behavioral constraints that guide the agent to operate safely within intended, user-controlled scopes.

    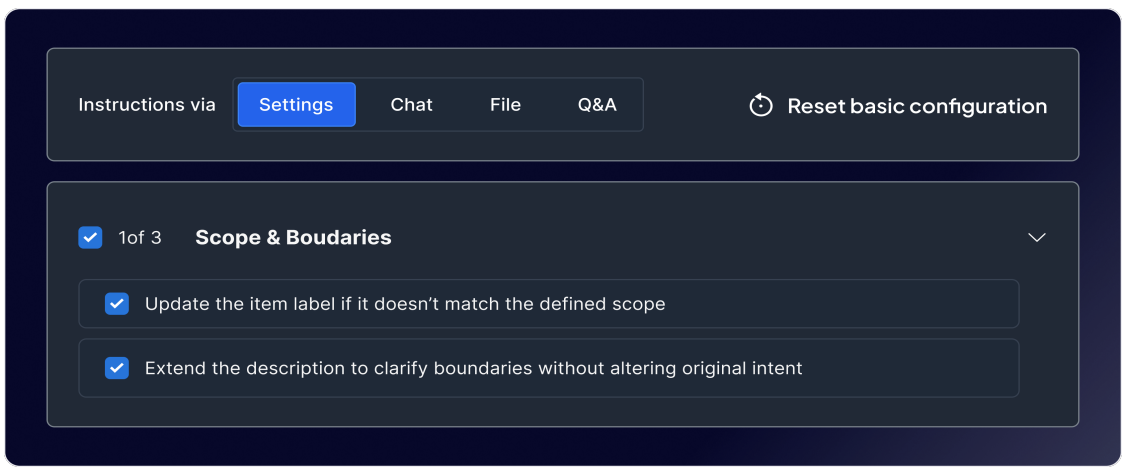

    Figure 1.

2. **Clarity:** Clarity involves making agent reasoning and decision processes legible, moving beyond black-box interactions toward interpretable, inspectable behavior. UI patterns include (not exhaustive) exposing inline rationales, confidence and uncertainty displays, source attributions, and alternatives or trade-offs, systems help users understand how outputs are produced, assess reliability, and decide when to accept, question, or adjust agent contributions.

    Figure 2 demonstrates this through confidence displays paired with inline rationale enabling users to evaluate both the system's logic and its level of certainty.

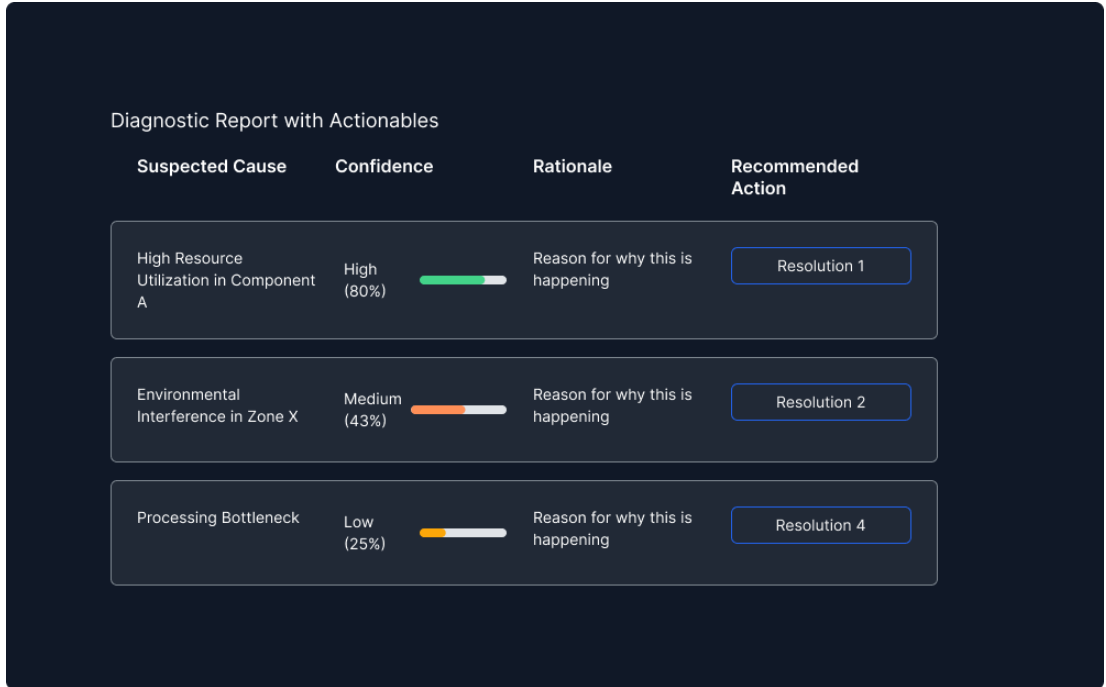

Figure 2.

3. **Recovery:** Recovery addresses the inevitability of error in probabilistic systems by ensuring that mistakes are tractable and reversible. Interfaces provide undo and redo support, editable outputs, safe defaults, and escalation paths to help users diagnose issues, correct course, and guide future behavior. These mechanisms transform breakdowns into opportunities for alignment and learning, maintaining trust and workflow momentum.

Figure 3 demonstrates undo and redo support: each automated change includes options to undo or revert, paired with a short, plain-language explanation of the change. Users can revert or approve individual actions or apply recovery to all at once, supporting both fine-grained control and efficient bulk handling.

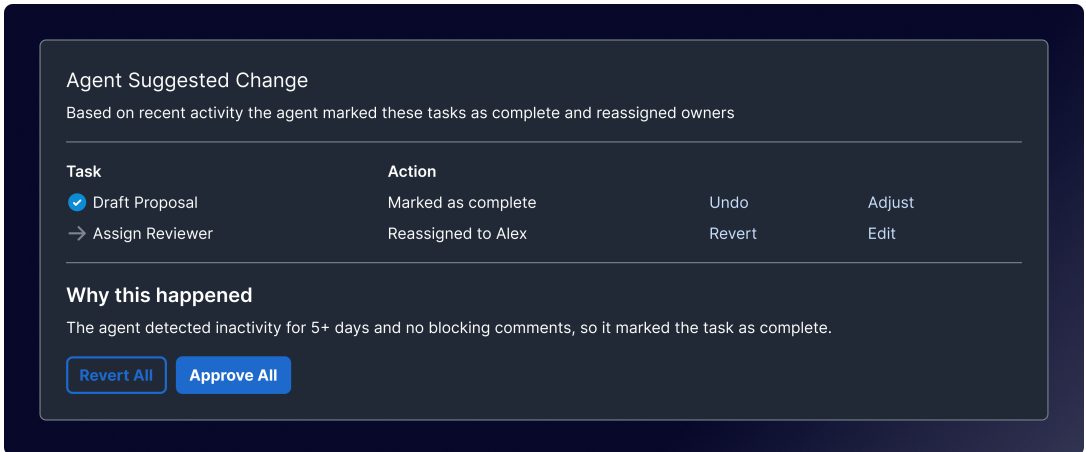

Figure 3.

4. **Collaboration:** Collaboration frames agents as active partners in mixed-initiative workflows, engaging in reciprocal interaction and shared ownership of outcomes. Through co-editing interfaces, role clarity and turn signals, and dynamic initiative management, systems support fluid transitions between human and agent contributions, leveraging complementary strengths to produce richer and more adaptive results.

Figure 4 demonstrates mixed-initiative interaction: the agent proactively detects issues or makes suggestions when it has useful context and can then refine its outputs in response to human edits or questions, keeping the workflow continuous and collaborative.

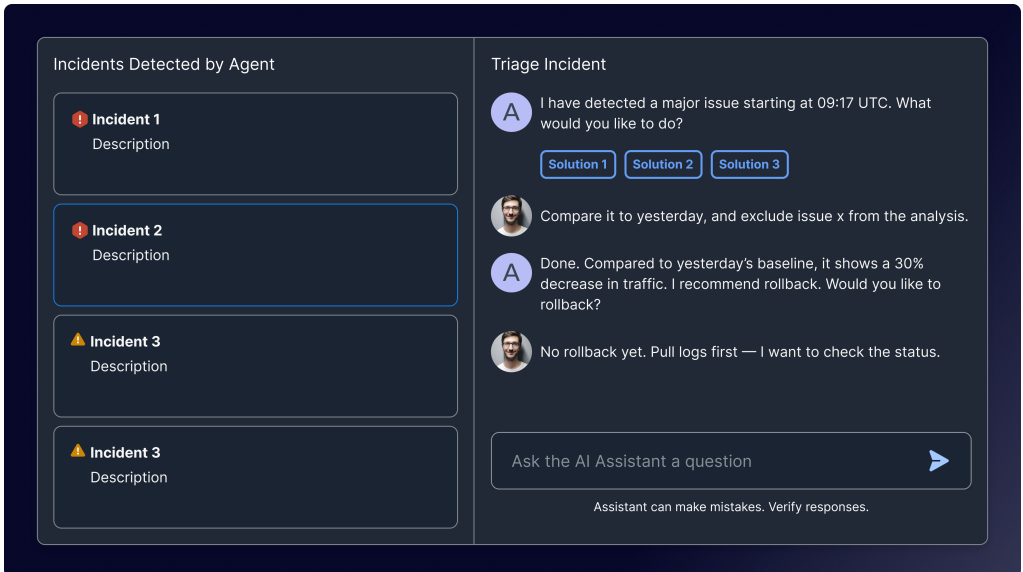

Figure 4.

5. **Traceability:** Traceability ensures that agent actions are temporally and contextually accountable, enabling retrospective understanding. Mechanisms such as action histories, visual diffing, and behavior tuning over time allow users and teams to inspect, audit, and refine decisions across sessions and actors, supporting transparency, debugging, and coordination in evolving multi-agent environments.

Figure 5 illustrates a log interface that lists all system and human actions in sequence with timestamps, shows how each step led to the next, records both user inputs and system decisions, and presents entries in clear, readable language, ensuring accessibility and accountability for all stakeholders.

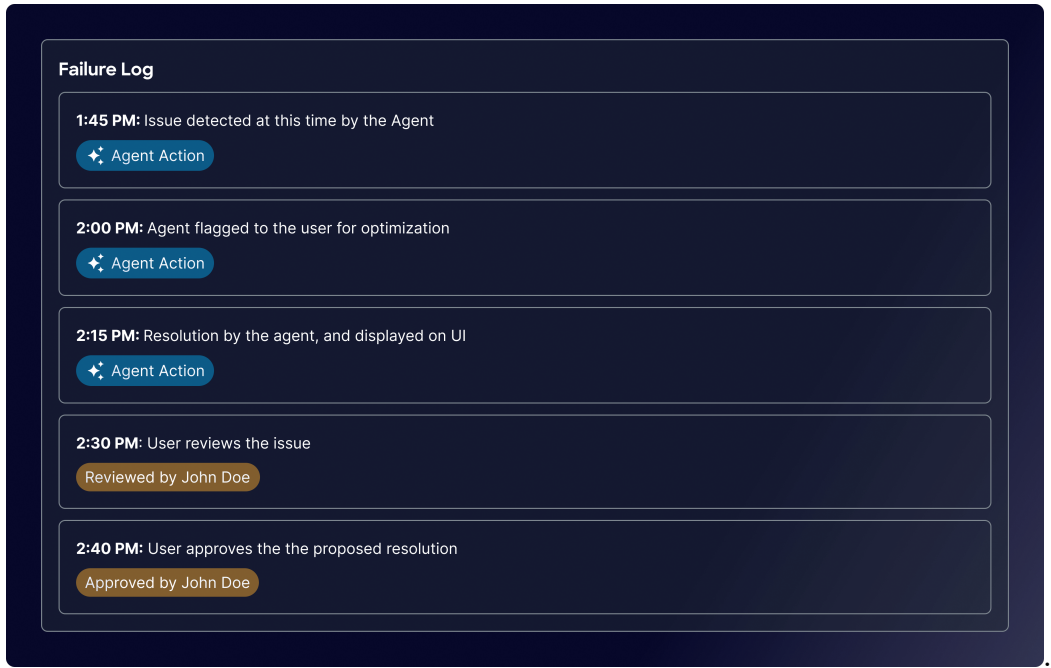

Figure 5.

---

## 5. The Agentic SDK: Schema-Driven Design

The HAX SDK operationalizes the principles outlined above by leveraging existing tool call functionalities in AI models to enforce predictable, structured agent outputs while preserving developer control over presentation logic. Rather than allowing agents to generate arbitrary UI instructions or inject unvalidated markup, the SDK constrains agentic interaction through tool calls that define permissible data structures and interaction primitives.

This design decision addresses a core tension in agentic interfaces, balancing the need to grant agents sufficient expressive capacity to convey intent against the unpredictability and security risks associated with unconstrained output generation. By utilizing the native tool-calling capabilities present in modern language models, the SDK creates a contract between agent reasoning and interface rendering—one that maintains type safety, enables static validation, and supports automated auditing of agent behavior.

A defining characteristic of the HAX SDK is its architectural commitment to embedding components directly within the developer's codebase rather than distributing them as opaque, externally managed dependencies. This decision reflects a fundamental philosophy that agentic interfaces require contextual adaptation that cannot be predetermined by framework authors.

When components are installed via the CLI, their complete source code is copied into the project's source tree. This includes UI components, type definitions, action hooks, and agent instruction files. Developers can directly modify rendering logic, adjust styling, refine prompt templates, or extend component behavior. This approach prioritizes transparency and local control over the abstraction offered by external package management. Because components exist as editable source rather than compiled dependencies, organizations can audit interface code, adapt components to internal design systems, and version-control modifications alongside application logic. This transparency enables tracing the full execution path from agent output to rendered UI, supporting debugging, compliance verification, and trust calibration.

**Customization Architecture**

1. React components define how structured agent outputs are visually presented. Developers can modify layouts, styling systems, design tokens, or substitute component implementations while maintaining tool call compatibility.
2. Each component includes an agent instruction file (`description.md`) that encodes how AI agents should reason about and populate that component's structure. These files are editable, allowing developers to refine prompt engineering, adjust output constraints, or introduce domain-specific guidance.

This dual-layer model decouples agent reasoning from interface presentation, allowing independent iteration on either layer without breaking the contract between them.

---

## 6. Distribution, Tooling, and Organizational Adoption

While the SDK establishes the architectural foundation for agentic interfaces, practical adoption requires tooling and mechanisms for organizational reuse. The HAX Command-Line Interface (CLI) provides an installation workflow, support for custom component repositories, and infrastructure for enterprise standardization.

The CLI integrates agentic interface patterns into existing React or Next.js projects. Developers initialize HAX within their project (`hax init`), which scaffolds the directory structure, configuration files, and type definitions. Components are installed on-demand via commands such as `hax add artifact <name>` or `hax add composer <name>`, which retrieve component source code and copy it into the project's HAX directory.

This installation model contrasts with traditional package managers by placing full source code under developer control rather than linking to external dependencies. The workflow integrates with existing version control and CI/CD practices.

Organizations often require internal component libraries that encode proprietary design standards, domain-specific interaction patterns, or compliance requirements. The HAX CLI implements a multi-repository architecture that allows teams to define and prioritize custom component registries alongside the official catalog.

Developers can configure multiple repository sources—such as internal instances, partner libraries, or development branches—and specify fallback resolution order. When installing a component, the CLI searches repositories in priority sequence, enabling organizations to override official components with internal alternatives while retaining fallback to the public registry.

## 7. Toward the Internet of Agents

As agents become composable, autonomous, and cross-domain, the infrastructure to connect them must be **open and interoperable**. The **AGNTCY initiative** [16] proposes the Internet of Agents as a new digital substrate, comprising:

- **Agent directories** with discoverability and reputation metrics
- **Open schemas** for capability description and verification
- **Protocols** for secure, quantum-safe communication
- **Evaluation frameworks** for trustworthiness and performance

By aligning UI patterns, SDK schemas, and orchestration logic with this broader infrastructure, the HAX framework contributes to a **frontend standard** for an emerging backend paradigm.

## 8. Conclusion and Future Work

This paper introduces a comprehensive framework for designing agentic interfaces — combining heuristics, structured schemas, orchestration logic, and reusable patterns. As we enter the age of distributed agents, UI design must evolve to support shared intent, reversible action, and collaborative execution.

In doing so, the framework directly responds to the problem articulated above: the persistence of legacy metaphors of strict control and command in an era defined by autonomous, context-sensitive, and generative systems.

Whereas traditional interface paradigms presuppose a human-centered locus of initiative, agentic systems require a model of reciprocal coordination, in which the boundaries of action, interpretation, and responsibility are continuously negotiated but ultimately within the control of the user. This work therefore reframes the interface not as a visual layer atop computation, but as a behavioral protocol.

Future work includes:

- Formal evaluations of trust patterns in user studies
- Benchmarking agentic interactions across verticals
- Integration with agent evaluation protocols from AGNTCY
- Extending the HAX SDK with support for embodied and voice agents
- Advancing the development of the HAX Agent as an adaptive orchestration layer for scalable context-aware interaction

We invite collaborators across design, AI, and infrastructure to help shape a more transparent, trustworthy, and human-centered Internet of Agents. The evolving HAX framework, including its principles, SDK, and component library, is available at [hax.design](hax.design) for continued exploration and contribution.

# References

[1] Lee JD, See KA. Trust in automation: designing for appropriate reliance. Hum Factors. 2004 Spring;46(1):50-80. doi: 10.1518/hfes.46.1.50_30392. PMID: 15151155

[2] Card, S. K., Moran, T. P., and Newell, A. The Psychology of Human–Computer Interaction. Lawrence Erlbaum Associates, 1983.

[3] Schömbs, O., Brendel, A. B., Kolbe, L., Peukert, C., and Benlian, A. HCI challenges and opportunities in interactive multi-agent systems. In Proceedings of the 2025 CHI Conference on Human Factors in Computing Systems (2025), pp. 1–14.

[4] Siddharth Mehrotra, Chadha Degachi, Oleksandra Vereschak, Catholijn M. Jonker, and Myrthe L. Tielman. 2024. A Systematic Review on Fostering Appropriate Trust in Human-AI Interaction: Trends, Opportunities and Challenges. ACM J. Responsib. Comput. 1, 4, Article 26 (December 2024), 45 pages. https://doi.org/10.1145/3696449

[5] Gille, F., Jobin, A., and Ienca, M. What we talk about when we talk about trust: Theory of trust for ai in healthcare. Intelligence-Based Medicine 1 (2020), 100001.

[6] Bansal, G., Wu, T., Zhou, J., Fok, R., Nushi, B., Kamar, E., Ribeiro, M. T., and Weld, D. Does the whole exceed its parts? the effect of ai explanations on complementary team performance. In Proceedings of the 2021 CHI Conference on Human Factors in Computing Systems (2021), pp. 1–16.

[7] Ma, S., Lei, Y., Wang, X., Zheng, C., Shi, C., Yin, M., and Ma, X. Who should i trust: Ai or myself? leveraging human and ai correctness likelihood to promote appropriate trust in ai-assisted decision-making. In Proceedings of the 2023 CHI Conference on Human Factors in Computing Systems
(New York, NY, USA, Apr 2023), CHI '23, Association for Computing Machinery, p. 1–19.

[8] Ahn, D., Almaatouq, A., Gulabani, M., & Hosanagar, K. (2024). Impact of model interpretability and outcome feedback on trust in AI. In Proceedings of the 2024 CHI Conference on Human Factors in Computing Systems (CHI '24) (Article 27, pp. 1–25). Association for Computing Machinery. https://doi.org/10.1145/3613904.3642780

[9] Chen, J., Mishler, S., and Hu, B. Automation error type and methods of communicating automation reliability affect trust and performance: An empirical study in the cyber domain. IEEE Transactions on Human-Machine Systems 51, 5 (2021), 463–473.

[10] Okamura, K., and Yamada, S. Adaptive trust calibration for human-ai collaboration. Plos one 15, 2 (2020), e0229132.

[11] Tomsett, R., Preece, A., Braines, D., Cerutti, F., Chakraborty, S., Srivastava, M., Pearson, G., and Kaplan, L. Rapid trust calibration through interpretable and uncertainty-aware ai. Patterns 1, 4 (2020), 100049.

[12] D. Gunning, M. Stefik, J. Choi, T. Miller, S. Stumpf, and G.-Z. Yang, "Xai—explainable artificial intelligence," Science robotics, vol. 4, no. 37, p. eaay7120, 2019.

[13] Zerilli, J., Bhatt, U., & Weller, A. (2022). How transparency modulates trust in artificial intelligence. Patterns, 3(4), 100455. https://doi.org/10.1016/j.patter.2022.100455

[14] Naikar, N., Brady, A., Moy, G., & Kwok, H. W. (2023). Designing human-AI systems for complex settings: ideas from distributed, joint, and self-organising perspectives of sociotechnical systems and cognitive work analysis. Ergonomics, 66(11), 1669–1694. https://doi.org/10.1080/00140139.2023.2281898

[15] McGrath, J. E. (1991). Time, interaction, and performance (TIP): A theory of groups. Small Group Research, 22(2), 147–174. https://doi.org/10.1177/1046496491222001

[16] Muscariello, L., Pandey, V., & Polic, R. (2025). The AGNTCY Agent Directory Service: Architecture and Implementation (arXiv preprint arXiv:2509.18787). https://doi.org/10.48550/arXiv.2509.18787